# A Serendipitous Deep Cluster Survey from ROSAT– PSPC pointed observations

P. ROSATI

*Department of Physics & Astronomy, The Johns Hopkins University, Baltimore, MD*

We present a deep X-ray selected sample of galaxy clusters which has been created from a serendipitous search in ROSAT-PSPC deep pointed observations at high galactic latitude. This survey, hereafter known as the ROSAT Deep Cluster Survey (RDCS), is being carried out utilizing a wavelet-based detection algorithm which, unlike other detection methods, is not biased against extended, low surface brightness sources. It is a flux-diameter limited sample that extends the X-ray flux limit of previous cluster surveys by more than one order of magnitude ($F_X \geq 1 \cdot 10^{-14} \mathrm{erg\,cm^{-2}s^{-1}}$). The first results of the on-going optical follow-up program indicate a high success rate of identification. At the present, 38 clusters out of 80 candidates have been identified on a 26 deg$^2$ surveyed area. Recently measured redshifts confirm the nature of these systems as low-moderate redshift groups ($z \simeq 0.2 - 0.3$) and intermediate to high redshift clusters ($z \simeq 0.4 - 0.7$). We show X-ray and optical images of several clusters identified to date, discuss the X-ray properties of the sample and present preliminary results on the redshift distribution. The final sample will include ∼100 clusters covering and area of ∼40 deg$^2$.

## 1. INTRODUCTION

X-ray surveys of clusters of galaxies provide ideal samples for studying the large scale structure and evolution in the Universe. Optically-selected cluster samples have commonly been used to constrain cosmological scenarios via studies of the correlation function and the volume distribution (e.g. Bahcall & Cen, 1992; Scaramella et al. 1991). In the last ten years however, the results of X–ray surveys have led to independent and often conflicting constraints casting doubts on the adequacy of optical catalogues. Indeed, optical surveys are well-known to have serious statistical shortcomings, such as contamination by chance coincidences of unvirialized systems and biases towards increasingly rich systems in a poorly understood, redshift-dependent manner. These difficulties can be significantly alleviated by selecting clusters via their X–ray emission which provides a unique tracer of cluster potential wells and is directly related to physical parameters of the cluster (dynamical state, density, temperature of the ICM). It should be kept in mind however that clusters in the X-ray sky come in a large variety of morphologies and apparent sizes as a result of their long dynamical time-scales, varying masses and temperatures. Thus, X–ray selection is still subject to certain biases, but these biases can be minimized and the statistical limits quantified *when an appropriate detection technique is used to identify extended X–ray sources and estimate their structural parameters.*

The ROSAT satellite is having a large impact on studies of galaxy clusters. The higher sensitivity and spatial resolution of the PSPC detector compared with previous X–ray missions provide an excellent opportunity to construct large cluster samples, deep enough to probe cosmological distances. Several X–ray cluster surveys at increasingly deep flux limits are currently underway using both survey mode data (e.g. Ebeling et al., 1994; Burg et al. 1994) and pointed observations (Rosati et al., 1995a,b; Castander et al., 1995). The final samples will constitute invaluable datasets for a rigourous study of the cluster X–ray luminosity function (XLF), the main tool for testing cosmological models. Two independent sets of X–ray observations have found negative evolution of the XLF at low-moderate redshifts (Edge et al. 1990; Henry et al. 1992), however these studies explored only a limited portion of the $L_X - z$ plane. The Edge et al. finding of strong evolution at a redshift of $\sim 0.1$ has recently been challenged by a bright All-Sky Survey (RASS) sample of Abell clusters (Ebeling et al. 1994). Furthermore, the evidence of steepening of the XLF from the Einstein Medium Sensitivity Survey (EMSS) cluster sample in the interval $0.15 \leq z \leq 0.6$ has a significance of $\sim 3\sigma$ (neglecting uncertainties other than number count statistics).

To date, the lack of a proper detection algorithm has been the major obstacle in compiling *pure* X–ray selected samples using RASS data (which are even hampered by lower spatial resolution) and only hybrid X–ray flux limited cluster samples have been constructed so far. These include a) *optically selected*, X–ray confirmed samples (Ebeling, 1994, at low redshifts; Bower et al., 1994, at higher redshifts) which rely on the correlation between $L_X$ and optical richness; b) 'mixed samples' containing extended X–ray sources and other 'point-like' sources which are correlated with galaxy overdensities in optical catalogues (Guzzo et al., 1994; Crawford et al. 1995). It is difficult to quantify the statistical completeness and establish the homogeneity of these samples, especially at high redshifts.

2. THE ROSAT DEEP CLUSTER SURVEY

In this spirit, we initiated a serendipitous search for galaxy clusters selected solely on the basis of their X–ray properties using deep ($T > 15$ ksec) high galactic latitude ROSAT–PSPC pointed observations drawn from the public archive. The RDCS is being carried out using a new source detection technique based on the Wavelet Transform which was developed specifically to carry out this project (Rosati, 1993, 1995a). This approach has the unique capability to detect and characterize faint X–ray sources with greater sensitivity over a wider range of sizes, morphologies and surface brightnesses than standard detection methods, thus probing a larger range of cluster parameters.

To date, we have processed 130 ROSAT-PSPC fields (covering $\sim 26$ deg$^2$) in a fully automatic way yielding 80 cluster candidates with the following criteria: **1-** $F_X[0.5-2.0 \text{ keV}] > 1 \cdot 10^{-14}$ erg cm$^{-2}$ s$^{-1}$; **2-** object extent exceeds the PSPC PSF by more than $3\sigma$; **3-** off-axis angle $< 15$ arcmin. In this process, several thousand point-like sources are used as a control sample to assess the statistical significance for a source to be extended. The sensitivity of the detection is modelled as a function of both the flux limit and the apparent size of the cluster candidates providing a rigorous estimate of the sky coverage and its uncertainties, essential for deriving the cluster number counts and the XLF (Rosati et al., 1995b (R95)).

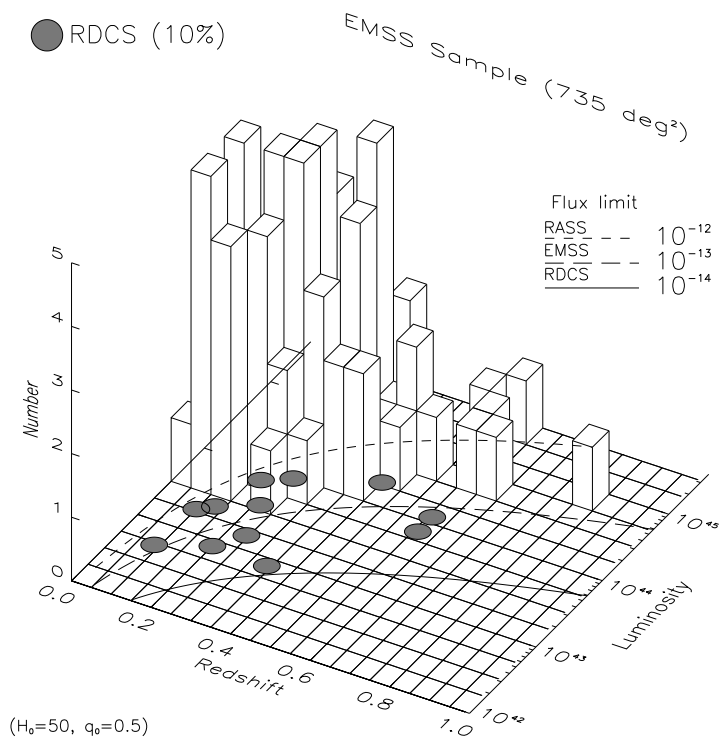

FIGURE 1. Distribution of the EMSS and RDCS clusters (with known redshifts) and respective flux limits in the redshift–luminosity plane. The X-ray luminosity is k-corrected and calculated in the [0.3–3.5] keV band.

The possible incompleteness of our sample due to misclassification of sources unresolved with the PSPC can be estimated either by modelling the X-ray surface brightness profile as a function of redshift and background, or cross-correlating the RDCS sample with several ROSAT deep surveys, whose optical identifications are now almost complete. Both these methods indicate that the incompleteness caused by the cut imposed on source extent is less than 15% at our flux limit (R95).

## 3. RESULTS FROM THE OPTICAL FOLLOW-UP PROGRAM

Inspection of digitised POSS-E plates show that the optical counterparts of these objects are too faint to be readily identified. In 1994 we began an optical follow-up program to identify these extended X–ray sources, via multi-band photometry and spectroscopy, at KPNO, CTIO and ESO. The first results of this program indicate an *extremely high success rate of identification*. At least 90% of the CCD images show a visible enhancement in the projected galaxy density at the peak of the X-ray emission. Of the 40 candidates imaged so far, 34 can be confidently classified as galaxy clusters (Table 1). Furthermore, deep imaging of three more candidates which lie *below* the completeness flux limit of $1 \cdot 10^{-14}$ erg cm$^{-2}$ s$^{-1}$ also indicates positive detections.

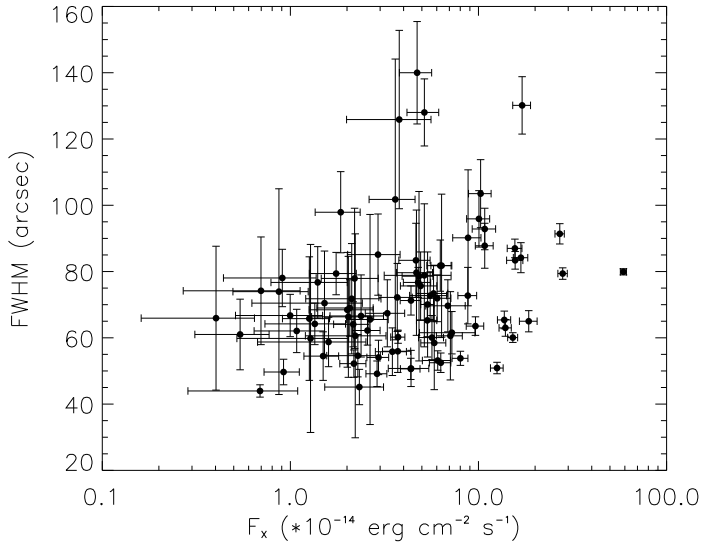

FIGURE 2. The flux–apparent size distribution of the RDCS cluster sample.

TABLE 1: RDCS Project Status (as of Sept 1995)

| | |
|---|---:|
| Processed X-ray fields | 130 |
| Surveyed area (deg$^2$) | 26 |
| Cluster candidates | 80 |
| Serendipitous rediscoveries | 4 |
| Imaged cluster candidates | 40 |
| Newly discovered clusters | 34 |
| Spurious | 1 |
| Unknown | 5 |
| Measured redshifts | 10 |

Recently measured redshifts confirm the nature of these systems as low–moderate redshift groups ($z \simeq 0.2 - 0.3$) and intermediate to high redshift clusters ($z \simeq 0.4 - 0.7$). In Fig.1 we show the bi-dimensional $L_X - z$ distribution of the EMSS sample and the RDCS clusters with known redshifts. It illustrates how the RDCS will probe a region of redshift–luminosity space mostly unexplored by previous, much shallower X-ray surveys. CCD images and overlaid X-ray contours for six clusters in our sample are shown in Fig.3. They are displayed in order of measured/estimated redshift representing systems at $z \sim 0.1$ (top), $z \simeq 0.3 - 0.5$ (middle) and $z > 0.5$ (bottom).

With an optically confirmed subsample, we have made the first measurement of the surface density of galaxy clusters at very low X–ray fluxes (7–10 clusters/deg$^2$ down to $1 \cdot 10^{-14}$ erg cm$^{-2}$ s$^{-1}$) (R95). At present, small number statistics and uncertainties in the low-luminosity end of the XLF prevent us from ruling out a no-evolution model using solely cluster number counts. In Fig.2 we plot the estimated FWHM of cluster candidates versus the X–ray flux in the band [0.5–2] keV. The median flux of the sample is $\sim 3 \cdot 10^{-14}$ erg cm$^{-2}$ s$^{-1}$.

TABLE 2: X–ray surveys sky coverage, flux limit and sample size

| Survey | $\Omega(\deg^2)$ | $f_{lim}$ [0.5–2] | N |
|---|---|---|---|
| RASS | $2.7 \cdot 10^3$ | $\sim 1 \cdot 10^{-12}$ | $\sim 1500$ |
| EMSS | 780 | $\sim 1 \cdot 10^{-13}$ | 104 |
| RIXOS | 15 | $3 \cdot 10^{-14}$ | 13 |
| RDCS | 40 | $1 \cdot 10^{-14}$ | $\sim 100$ |

When the remaining 25% of the archive is processed, we expect the final catalogue will contain more than 100 clusters over an area of $\sim 40$ deg$^2$ (see Table 2), ranging from low X–ray luminosity systems to high luminosity, rich clusters with redshift of order of unity. The completion of the redshift survey will allow us to constrain the unknown faint end of the cluster XLF at moderate-high redshifts. In addition, the RDCS will complement the results of shallower, larger surveys, such as the RASS which probes the XLF on a wider range of luminosities and a lower redshifts. The combination of the RASS and RDCS samples will ultimately place tight constraints on cluster evolution over a substantial fraction of a Hubble time and will constitute a fundamental dataset with which to test cosmological models.

ACKNOWLEDGEMENTS

I would like to thank my collaborators Richard Burg, Riccardo Giacconi, Roberto Della Ceca, Colin Norman and Brian McLean who contributed to the work presented here. I acknowledge partial support from NASA grants NAG5-1538, NAGW-2508, NAG8-794.

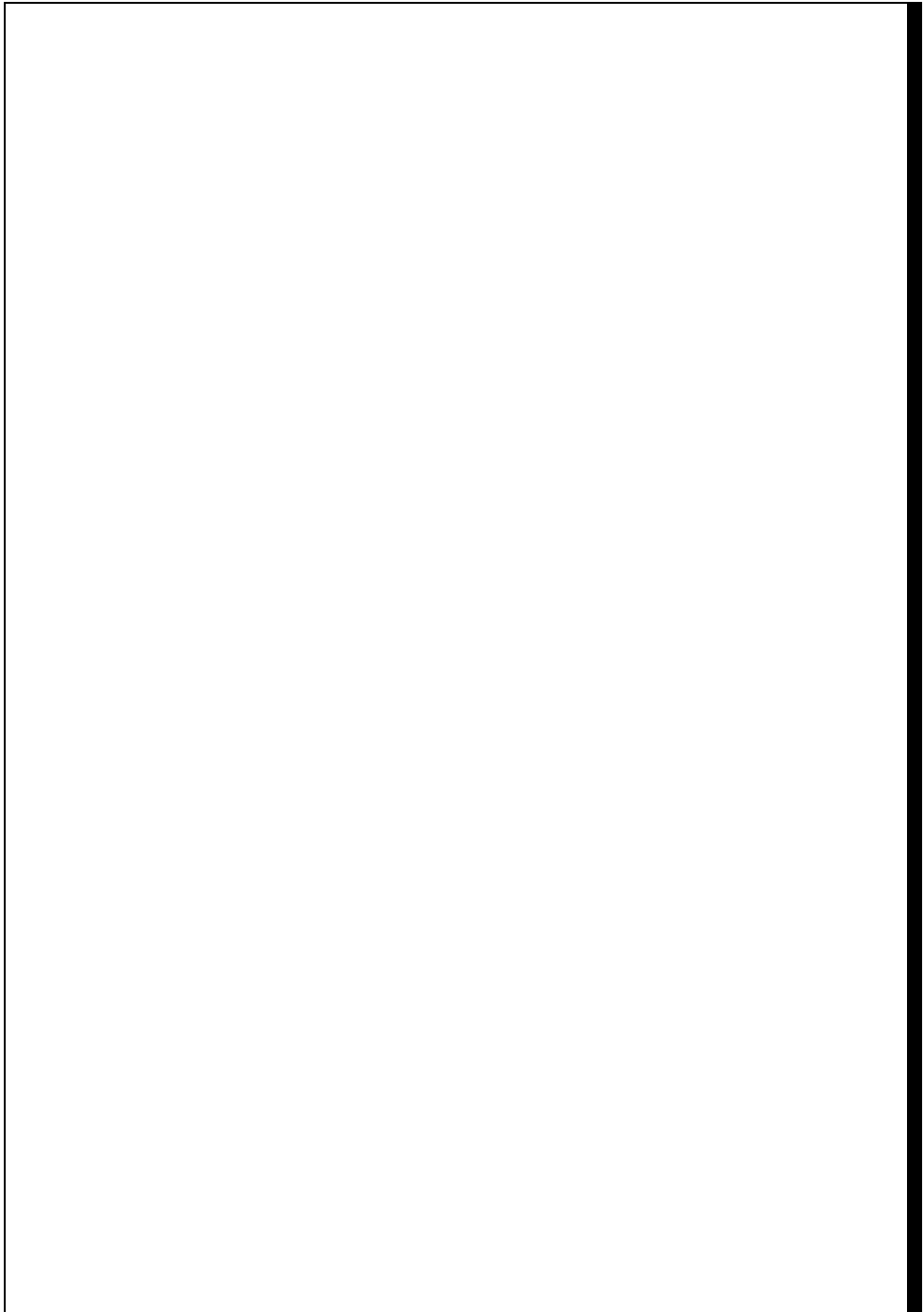

FIGURE 3. ROSAT PSPC X-ray contours ([0.5–2] keV band) ovelaid on CCD R/I-band images for several identified groups/clusters of the RDCS sample. The images are 3 arcmin across; the contours are at [3,5,7,10]$\sigma$ above the background.

6